\documentclass{PoS}

\def\be{\begin{equation}}
\def\ee{\end{equation}}
\def\bea{\begin{eqnarray}}
\def\eea{\end{eqnarray}}
\def\ms{\overline{MS}}
\def\ov{\overline}
\def\t{\tilde }
\def\Lam{\Lambda}

\def\ga{\gamma}

\title{A new determination of $\alpha_S$ from Renormalization Group Optimized Perturbation}

\ShortTitle{$\alpha_S$ from $F_\pi$ and Renormalization Group Optimized Perturbation}

\author{\speaker{Jean-Lo\"{\i}c KNEUR} and Andr\'e NEVEU\\
       CNRS, Laboratoire Charles Coulomb UMR 5221, F-34095, Montpellier, France\\
Universit\'e Montpellier 2, Laboratoire Charles Coulomb UMR 5221, F-34095, Montpellier, France\\
        E-mail: \email{jlkneur@univ-montp2.fr}, \email{andre.neveu@univ-montp2.fr}  }

\abstract{A new version of the so-called optimized perturbation (OPT), 
 implementing consistently renormalization group properties, 
is used to calculate the nonperturbative ratio $F_\pi/\Lam_{\ms}$ of 
the pion decay constant and the basic QCD scale in the $\ms$ scheme. Using the experimental $F_\pi$ input value it provides 
a new determination of $\Lam_{\ms}$ for $n_f=2$ and $n_f=3$, 
and of the QCD coupling constant  $\overline\alpha_S $ at various scales once combined with a standard 
perturbative evolution. The stability and empirical convergence properties of the RGOPT modified series is demonstrated up to the third 
order. We examine the difference sources of theoretical uncertainties and 
obtain $\overline\alpha_S (m_Z) =0.1174 ^{+.0010}_{-.0005} \pm .001 \pm .0005_{evol}$, where the first errors are 
estimates of the intrinsic theoretical uncertainties of our method, and the 
second errors come from present uncertainties in $F_\pi/F_0$, where $F_0$ is $F_\pi$ in the exact chiral $SU(3)$ limit.}

\FullConference{The European Physical Society Conference on High Energy Physics \\
    18-24 July, 2013\\
  Stockholm, Sweden}

\begin{document}

\section{Introduction and Motivation}
In the chiral symmetric, massless quark  limit, the QCD coupling $\ov\alpha_S(\mu)$ or equivalently  
\be
\ov\Lam^{n_f} \sim \mu\, e^{-\frac{1}{\beta_0 \alpha_S}}(1+\cdots)
\label{Lam}
\ee
in a specified renormalization scheme, is the only QCD parameter~\footnote{In Eq.~(\ref{Lam}) ellipses stand for  
higher RG orders corrections and $\Lambda^{n_f}_{\ms}$ depends also on the number of active quark flavors $n_f$,  
with perturbative matching at the quark mass thresholds\cite{PDG}}.  
The present world average value~\cite{PDG}  $\ov\alpha_S(m_Z)=.1184 \pm .0007$  is impressively accurate, 
combining many different determinations.
It is worth however to obtain $\ov\Lam$ or $\ov\alpha_S(\mu)$ from further independent analyses and methods, 
specially for $n_f=2$ in the infrared range not perturbatively extrapolable from high scale values.\\ 
Our more general purpose is to obtain approximations of reasonable accuracy to the chiral symmetry breaking order parameters, $F_\pi$, $\langle \bar q q \rangle$, etc, 
from an alternative (optimized) use of perturbation series, exploiting that
typically the pion decay constant $F_\pi$ should be entirely 
determined by $\ov\Lam$ in the strict chiral limit, from which $\alpha_S$ can be obtained as a by-product. 
However the intrinsically nonperturbative ratio $F_\pi/\ov\Lam$ is normally not accessible from standard perturbative calculations: 
first, naively it should involve a priori large $\alpha_S(\mu)$ values at the presumably corresponding low scales $\mu\sim\Lam$, 
invalidating ordinary QCD perturbative expansions. 
Moreover, an often invoked argument is that, even if $F_\pi$ and  $\langle \bar q q \rangle$ have standard 
perturbative QCD expansions, these are proportional to the light quark masses, e.g. $F^2_\pi \sim m^2_q 
\sum_{n,p} \alpha^n_s\ln^p (m_q)$, thus trivially vanishing
anyway in the relevant chiral limit $m_q\to 0$ at arbitrary perturbative orders. 
As we will see, one can in fact circumvent both problems, by an appropriate  
modification of the ordinary perturbative expansion.
\section{ Optimized Perturbation (OPT)}
The first basic idea~\cite{delta} is to 
reshuffle the standard QCD Lagrangian by introducing 
an extra parameter $0<\delta<1$, interpolating between ${\cal L}_{free}$ and 
${\cal L}_{int}$, such that the mass 
 $m_q$ is traded for an arbitrary ``trial'' parameter.
This is perturbatively equivalent
to taking standard perturbative expansions in $g\equiv 4\pi\alpha_S$, after renormalization, reexpanded  
in powers of $\delta$ after substituting:
\be m_q \to  m\:(1- \delta)^a,\;\; g \to  \delta \:g\;.
\label{subst1}
\ee
The whole procedure is consistent with renormalizability~\cite{gn2,qcd1} and gauge invariance~\cite{qcd1}. 
In (\ref{subst1}) we introduce an extra parameter $a$, to reflect a priori a certain freedom 
in the interpolating form, allowing to impose further physical 
constraints as will be discussed below.  
Applying (\ref{subst1}) to a given perturbative expansion for a physical quantity $P(m,g)$, reexpanded in $\delta$ 
at order $k$, and taking {\em afterwards} the $\delta\to 1$ limit to recover the original {\em massless} theory,  
leaves a remnant $m$-dependence at any finite $\delta^k$-order. 
Then $m$ is most conveniently fixed by an optimization (OPT) prescription:
\be
\frac{\partial}{\partial\,m} P^{(k)}(m,g,\delta=1)\vert_{m\equiv \tilde m} \equiv 0\;,
\label{OPT}
\ee  
thus determining a nontrivial optimized mass $\t m(g)$, of order $\Lam$, realizing dimensional transmutation, unlike the original mass vanishing in the chiral limit.  
But does this 'cheap trick' always work? or when (and why) does it work?
In simpler ($D=1$) models such a procedure may be seen as a particular case of 
``order-dependent mapping''\cite{odm},  
which has been proven\cite{deltaconv} to converge exponentially fast for the $D=1$ $\Phi^4$ oscillator energy levels.  
For $D > 1$ renormalizable models, the large $\delta$-orders behaviour  
is more involved, and no equivalent convergence proof exists at present (although OPT is seen to partially  
damp the factorially divergent (infrared renormalons) perturbative behaviour at large orders~\cite{Bconv}).   
Nevertheless the OPT can give rather successful   
approximations to certain nonperturbative quantities beyond mean field approximations in a wide variety of models~\cite{delta,beccrit,bec2}, also in studies of 
phase transitions at finite temperatures and densities.
\section{Renormalization Group improvement of OPT (RGOPT)} 
Most previous OPT applications are based on the so-called linear $\delta$-expansion, assuming $a=1$ 
in Eq.~(\ref{subst1}) mainly for simplicity and economy of parameters. 
Our more recent approach\cite{rgopt1,rgopt_Lam,rgopt_alphas} differs in two respects, which turn out to drastically improve the convergence.
First, we introduce a straightforward marriage of OPT and renormalization group (RG)  
properties, by  
requiring the ($\delta$-modified) expansion to satisfy, in addition to the OPT Eq.~(\ref{OPT}), a 
standard RG equation:
\be
\mu\frac{d}{d\,\mu} \left(P^{(k)}(m,g,\delta=1)\right) =0, 
\label{RG}
\ee 
where the (homogeneous) RG operator acting on a physical quantity is\footnote{Our normalization is $\beta(g)\equiv dg/d\ln\mu = -2b_0 g^2 -2b_1 g^3 +\cdots$,  
$\gamma_m(g) = \gamma_0 g +\gamma_1 g^2 +\cdots$ with $b_i$, $\gamma_i$ up 
to 4-loop given in~\cite{bgam4loop}.} $\mu\frac{d}{d\,\mu} =
\mu\frac{\partial}{\partial\mu}+\beta(g)\frac{\partial}{\partial g}-\gamma_m(g)\,m
 \frac{\partial}{\partial m}$. 
Since interaction and free terms from the original perturbative series are reshuffled by the $\delta$-modified
expansion, Eq.~(\ref{RG}) gives a nontrivial additional constraint.  Moreover, 
combined with Eq.~(\ref{OPT}), the RG equation takes a reduced form:
\be
\left[\mu\frac{\partial}{\partial\mu}+\beta(g)\frac{\partial}{\partial g}\right]P^{(k)}(m,g,\delta=1)=0\;.
\label{RGred}
\ee
As a result Eqs.~(\ref{RGred}) and (\ref{OPT}) together completely fix 
{\em optimized} $m\equiv \t m$ and $g\equiv \t g$ values.\\
Now a well-known drawback of the standard OPT approach is that, beyond lowest order, Eq.~(\ref{OPT}) generally gives more and more solutions 
at increasing orders, furthermore many being complex. Without more insight on the nonpertubative behaviour, it may be difficult
to select the right solution, and the unphysical complex solutions are embarrassing. 
But RG considerations also provide possible ways out, and this is the second main difference and new feature of our RGOPT version.  
More precisely for QCD (more generally for any asymptotically free (AF) models), we proposed~\cite{rgopt_Lam,rgopt_alphas} 
a rather compelling selection criterion, retaining only the solution(s) continuously matching asymptotically the standard AF perturbative RG behaviour for $g\to 0$: 
\be
\t g (\mu \gg \t m) \sim (2b_0 \ln \frac{\mu}{\t m})^{-1} +{\cal O}( (\ln \frac{\mu}{\t m})^{-2})\;.
\label{rgasympt}
\ee
This provides a unique solution at a given order for both the RG and OPT equations. A welcome additional feature is that
requiring at least one RG solution to fulfill (\ref{rgasympt}) gives in fact a strong necessary condition on the basic interpolation (\ref{subst1}), fixing $a$ uniquely
in terms of the universal (scheme-independent) first order RG coefficients: 
\be
a\equiv \ga_0/(2b_0)\;.
\label{acrit}
\ee
A connection of $a$ with RG anomalous dimensions/critical exponents 
was also established in a very different context, in the  $D=3$ $\Phi^4$ model for the Bose-Einstein condensate (BEC) 
critical temperature shift by two independent OPT approaches~\cite{beccrit,bec2}, where
it also led to real OPT solutions~\cite{bec2}.  
Unfortunately, in the more involved QCD-like theories, AF-compatibility and reality of solutions appear in general mutually exclusive beyond lowest order. 
A simple way out is to further exploit the RG freedom, by considering a perturbative renormalization scheme change to possibly 
recover  RGOPT solutions both AF-compatible and real~\cite{rgopt_alphas}, as we discuss below. 
\section{Applications: $F_\pi/\Lambda_{\ms}$}
As an order parameter of the  
 chiral symmetry breaking $SU(n_f)_L\times SU(n_f)_R \to SU(n_f)_{L+R}$ 
for $n_f=2$ or $n_f=3$ massless quarks, $F_\pi$ is related to the $p^2\to 0$ axial current correlator:
\be
i \langle 0| T A^i_\mu (p) A^j_\nu (0)|0 \rangle \equiv \delta^{ij}
 g_{\mu\nu} F^2 +{\cal O}(p_\mu p_\nu)\label{Fpidef}
\ee
where $A^i_\mu \equiv \bar q \gamma_\mu \gamma_5 \frac{\tau_i}{2} \:q$,
and $F$ is $F_\pi$ in the strict chiral limit.
The perturbative expansion of (\ref{Fpidef}) in the $\ms$ scheme, with quark masses $m\ne 0$, can be extracted from known 4-loop
calculations~\cite{Fpi_4loop}: 
\be
F^2_\pi(m) = 3 \frac{m^2}{2\pi^2} \left[ -L +\frac{\alpha_S}{4\pi}(8 L^2
+\frac{4}{3} L +\frac{1}{6}) 
+(\frac{\alpha_S}{4\pi})^2 [f_{30} L^3+f_{31} L^2 +f_{32}L +f_{33}]+{\cal O}(\alpha^3_S)
\right]  +\mbox{div}
\label{Fpipert}
\ee    
where $L\equiv \ln \frac{m}{\mu}$, and the coefficients $f_{ij}$ 
are given in \cite{rgopt_alphas}. 
Prior to performing the OPT, one subtlety is that in dimensional regularization
(\ref{Fpipert}) requires an extra (subtraction) renormalization:
\be
F^2_\pi(\mbox{RG-inv})\equiv  F^2_\pi -S(m,\alpha_S);\;\;\;
S(m,\alpha_S)\equiv m^2 (s_0/\alpha_S + s_1 +s_2 \alpha_S +...)
\label{sub}
\ee
on dimensional grounds, due to composite operator mixing. 
To obtain a finite {\em and} RG-invariant expression, this subtraction should be performed 
consistently with RG properties, and the perturbative coefficients $s_k$ in (\ref{sub}) are fixed  
from the coefficient of the $L$ term at order $k+1$~\cite{qcd1,rgopt_alphas}.  One finds 
$s_0 = 3/(16\pi^3)(b_0-\gamma_0)$, etc\cite{rgopt_alphas}.\\

We can now apply to the (subtracted) RG-invariant perturbative series for $F$ the procedure (\ref{subst1}), 
at orders $\delta^k$, then solving OPT and RG Eqs.~(\ref{OPT}), (\ref{RGred}). 
In the first RG order approximation, and neglecting non-logarithmic terms at all orders in (\ref{Fpipert}), results are exact and very transparent:
the RG and OPT Eqs.~(\ref{RGred}),(\ref{OPT}) have a unique, AF-compatible, real solution:
\be
 \t L \equiv \ln \frac{\t m}{\mu} =-\frac{1}{2b_0 g}=-\frac{\gamma_0}{2b_0}\;;\;\;\;\;\t \alpha_S = \frac{1}{4\pi\ga_0} =\frac{\pi}{2}
 \ee
which already gives a quite realistic value $F(\t m,\t \alpha_S)= (\frac{5}{8\pi^2})^{1/2} \t m \simeq  0.25  \Lambda_{\ms}$. 
Including higher RG and non-RG order terms, 
the RG and OPT equations, polynomial in $(L,g)$,
thus give at increasing $\delta$-orders 
(too) many solutions, most being complex (conjugates). But requiring the $L(g)$ solutions to
match the standard AF perturbative behaviour (\ref{rgasympt}) gives unique OPT and RG branches
and completely fix the critical value (\ref{acrit}). Unfortunately 
the intersection of these AF-compatible RG and OPT branches is complex at $\delta^k, k\ge 1$ orders in the $\ms$-scheme. 
\subsection{Recovering real AF-compatible solutions}
The non-reality of AF-compatible solutions is an artifact 
of solving exactly polynomial equations, and to some extent also an accident of the $\ms$-scheme. 
Thus one can expect to recover real solutions by a moderate (perturbative) deformation of the (AF-compatible) RGOPT solutions. 
A deformation consistent with RG properties is simply a standard 
perturbative renormalization scheme change (RSC). Since $m$ is a spurious parameter, we consider RSC affecting only $m$: 
\be
 m\to  m' (1+B_1 g+B_2 g^2+\cdots)
 \label{rsc}
 \ee
 which also let the RG Eq.~(\ref{RGred}) and $\Lam_{\ms}$ unaffected. One can now move in the RSC parameter space 
 with some $B_k\ne 0$ in (\ref{rsc})
 to possibly reach real solutions.
 Now to stay as near as possible to the reference $\ms$-scheme, we require a {\em closest contact} solution between the RG and OPT curves 
 (thus closest to $\ms$), analytically given by colinearity of the tangent vectors of the two curves:
 \be
\frac{\partial}{\partial g} \mbox{RG}(g,L,B_k) \frac{\partial}{\partial L} \mbox{OPT}(g,L,B_k) -\frac{\partial}{\partial L} \mbox{RG}\, 
\frac{\partial}{\partial g} \mbox{OPT} \equiv 0 
\ee
to be solved together with the OPT and RG Eqs.~(\ref{OPT}),(\ref{RGred}) now for $\t L, \t g, \t B_k$ at successive orders.  
From basic RSC properties the differences with respect to the original $\ms$-scheme are expected to decrease at higher perturbative order. 
Thus, besides recovering real solutions, RSC provides well-defined uncertainty estimates, since  
non-unique RSC prescriptions imply different results, that we take as 
intrinsic theoretical uncertainties of the method. The main results are shown for $n_f=3$ in Table \ref{tabrealn3} (normalized with a 4-loop
perturbative $\overline{\Lam}_{4l}$ expression (\ref{Lam})). Note that $\t m'/\ov\Lam$ is ${\cal O}(1)$ and $\t \alpha_S$ stabilizes to reasonably perturbative values $\simeq 0.4$. For the similar $n_f=2$ results~\cite{rgopt_alphas}, the theoretical RSC uncertainties
are about twice larger (which can be traced  to the larger RSC corrections and somewhat larger $\t \alpha_S$ needed to reach real solutions).  
\begin{table}[h!]
\begin{center}
\caption{Main optimized results at successive orders for $n_f=3$} 
\begin{tabular}{|l||c|c|c|c|}
\hline
$\delta^k$ order & nearest-to-$\ms$  RSC $ \t B_i $  &  $\t L'$ & $\t \alpha_S$ & $ \frac{F_0}{\overline\Lam_{4l}}$ (RSC uncertainties) \\
\hline
$\delta$, RG-2l & $\t B_2 =2.38 \, 10^{-4}$  &   $-0.523 $    & $0.757$ & $ 0.27-0.34$ \\
$\delta^2$,  RG-3l& $ \t B_3 =3.39\,10^{-5}$  & $ -1.368 $ & $ 0.507 $  & $ 0.236-0.255 $  \\
$\delta^3$, RG-4l &$\t B_4 =1.51\,10^{-5}$ & $-1.760$ & $0.374$ & $0.2409-0.2546$\\
\hline
\end{tabular}
\label{tabrealn3}
\end{center}
\end{table}
%
\subsection{Explicit symmetry breaking and extracting $\alpha_S(\mu)$}
Now to get a more realistic result we should remember that the above calculation is actually for $F_\pi(m_q\to 0)$ 
in the exact chiral limit, and ``subtract out'' the explicit chiral symmetry breaking effects from non-zero  
$m_u, m_d, m_s$, specially important in the $SU(3)$ case. Denoting $F$ and $F_0$ the $F_\pi$
values in the chiral $SU(2)$ and $SU(3)$ respectively, recently combined lattice results are~\cite{LattFLAG}:
\be
 \frac{F_\pi}{F} \sim 1.073 \pm 0.015; \;\;\;\;\;\;\;
 \frac{F_\pi}{F_0} \sim 1.172(3)(43) 
 \ee
where the $F$ value is robust, and $F_0$ was obtained by the MILC collaboration~\cite{milc10} using fits to next-to-next-leading order Chiral Perturbation~\cite{chptNNLO}. 
It is worth however to keep in mind that lower $F_0$ values are also advocated~\cite{LattFLAG}, hinting at a slower convergence of $n_f=3$ Chiral 
Perturbation~\cite{F0low,chptNNLO}. With the different theoretical uncertainties combined we thus obtain: 
\be
\overline\Lam^{n_f=2}_{4l} \simeq 359^{+38}_{-26} \pm 5\; \mbox{MeV};\;\;\;\overline\Lam^{n_f=3}_{4l} \simeq  317 ^{+14}_{-7} \pm 13\; \mbox{MeV}.
\label{Lam3}
\ee
This compares reasonably well 
with different kinds of lattice determinations for $n_f=2$ or $n_f=3$ (see \cite{rgopt_alphas} for a detailed comparison with Lattice $\overline\Lam$ results).
Finally using $\overline\Lam(n_f=3)$ from (\ref{Lam3}) and a standard perturbative evolution at 4-loop including $m_c$, $m_b$ threshold effects~\cite{matching4l}
we obtain
\be
\overline\alpha_S(m_Z) =0.1174 ^{+.0010}_{-.0005}|_{\mbox{th,rgopt}} \pm .0010_{\,\delta F_0}  \pm .0005_{\mbox{evol}}\;.
\label{alphas}
\ee
\section{Summary}
Our version of the OPT with consistent RG properties implies a basic interpolation (\ref{subst1}) with $a$ in (\ref{acrit}) $\ne 1$ uniquely determined by universal RG
coefficients. This appears to drastically improve the convergence and stability properties. Calculating $F_\pi/\overline{\Lam}$ at three successive orders
we extract rather precise $\overline \alpha_S$ values (\ref{alphas}), taking into account intrinsic theoretical uncertainties of the method.


\begin{thebibliography}{99}
\bibitem{PDG} S. Bethke,  G. Dissertori and G.~P.~Salam in
J.~Beringer {\it et al.} 
  Phys.\ Rev.\ D {\bf 86} (2012) 010001.
%
\bibitem{delta} There are numerous references on the delta-expansion, see
{\it e.g.} ref. [21] in \cite{rgopt_alphas}.  
%
\bibitem{gn2} C.~Arvanitis, F.~Geniet, M. Iacomi, J.-L.~Kneur and A.~Neveu,
Int. J. Mod. Phys. {\bf A12}, 3307 (1997).
%
\bibitem{qcd1}  J.-L. Kneur, Phys. Rev. {\bf D57}, 2785 (1998).
%
\bibitem{odm} R. Seznec and J. Zinn-Justin, J. Math. 
 Phys. {\bf 20}, 1398 (1979); J.C. Le Guillou and J. Zinn-Justin,
Ann. Phys. {\bf 147}, 57 (1983). 
%
\bibitem{deltaconv} R. Guida, K. Konishi and H. Suzuki, Ann. Phys. {\bf 241}
(1995) 152;  Ann. Phys. {\bf 249}, 109 (1996); 
%
\bibitem{Bconv} J.-L. Kneur and D. Reynaud, Phys. Rev. {\bf D66}, 085020 (2002).
%
\bibitem{beccrit} H.~Kleinert, Mod. Phys. Lett. {\bf B17}, 1011 (2003);
B.~Kastening, Phys. Rev. {\bf A68}, 061601 (2003); 
Phys.Rev. {\bf A69} (2004) 043613.
%
\bibitem{bec2} J.-L.~Kneur, A.~Neveu and M.B~Pinto,
Phys. Rev. {\bf A69} 053624 (2004).
\bibitem{rgopt1} J.-L.~Kneur and A.~Neveu, Phys. Rev. {\bf D81},~125012 (2010). 
%
\bibitem{rgopt_Lam}  J.~-L.~Kneur and A.~Neveu,
  Phys.\ Rev.\ D {\bf 85} (2012) 014005.
%
\bibitem{rgopt_alphas}  J.~-L.~Kneur and A.~Neveu,
  arXiv:1305.6910 [hep-ph], Phys.\ Rev.\ D {\bf 88} (2013) 074025.
%
\bibitem{bgam4loop} J.~A.~M.~Vermaseren, S.~A.~Larin and T.~van Ritbergen,
  Phys.\ Lett.\  B {\bf 405}, 327 (1997).
%
\bibitem{Fpi_4loop} K.~G.~Chetyrkin, J.~H.~Kuhn and M.~Steinhauser,
  Nucl.\ Phys.\ B {\bf 505} (1997) 40;
A.~Maier et al, Nucl.\ Phys.\  B {\bf 824} (2010) 1; and refs. therein.
%
\bibitem{LattFLAG}G.~Colangelo et al, Eur. Phys. J. {\bf C71}, 1695 (2011).
%
\bibitem{milc10} A.~Bazavov {\it et al.}  [MILC Collaboration],
  PoS CD {\bf 09} (2009) 007.
%
\bibitem{chptNNLO} J.~Bijnens and T.~A.~Lahde,
  Phys.\ Rev.\ D {\bf 71} (2005) 094502.
\bibitem{F0low} See e.g. S.~Descotes-Genon, N.~H.~Fuchs, L.~Girlanda and J.~Stern, Eur.\ Phys.\ J. C{\bf 34} (2004) 201.
%
\bibitem{matching4l} K.~G.~Chetyrkin, J.H K\"uhn and C.~Sturm,  Nucl.\ Phys.\ B {\bf 744} (2006) 121 and refs. therein.
\end{thebibliography}
\end{document}